\documentclass[pra,twocolumn,notitlepage,floatfix]{revtex4-2}
\usepackage[T1]{fontenc}
\usepackage[colorlinks=true,urlcolor=blue,citecolor=blue,anchorcolor=red]{hyperref}
\usepackage{graphics,url,color,physics,bbm,soul,mathtools,amssymb,amsmath,dsfont,booktabs}

\begin{document}

\title{Classically optimized Hamiltonian simulation}

\author{Conor Mc Keever}
\email{conor.mckeever@quantinuum.com}
\affiliation{Quantinuum, Partnership House, Carlisle Place, London SW1P 1BX, United Kingdom}

\author{Michael Lubasch}
\affiliation{Quantinuum, Partnership House, Carlisle Place, London SW1P 1BX, United Kingdom}

\date{May 10, 2023}

\begin{abstract}
Hamiltonian simulation is a promising application for quantum computers to achieve a quantum advantage.
We present classical algorithms based on tensor network methods to optimize quantum circuits for this task.
We show that, compared to Trotter product formulas, the classically optimized circuits can be orders of magnitude more accurate and significantly extend the total simulation time.
\end{abstract}

\maketitle

\section{Introduction}

Quantum computers are expected to produce novel insights into quantum chemistry and materials~\cite{KaEtAl11, CaEtAl19, McEtal20, BaEtAl20} as they are believed to have an advantage over classical computers in performing Hamiltonian simulation, since powerful quantum algorithms for this task were discovered~\cite{Wi96, Ll96, AbLl97, Za98, KaEtAl08}.
To make this type of simulation more efficient, there has been significant progress on the theoretical quantum algorithmic side~\cite{BeCh12, ChWi12, BeEtAl15, BeChKo15, LoCh17, LoCh19} as well as on the more applied side of variational quantum algorithms~\cite{LiBe17, YuEtAl19, cirstoiu2020variational, LiEtAl20, BarEtAl20, YaEtAl21, BeFiLu21, BaViCa21, WaEtAl21, MiEtAl22}.
However, the theoretical approaches can lead to deep quantum circuits and the variational approaches can require a large number of circuit runs in the optimization.
Therefore Hamiltonian simulation is still challenging for current quantum devices~\cite{SmEtAl19, FaZh20, GuEtAl19}.

In this article, our goal is to reduce the quantum hardware requirements for Hamiltonian simulation by performing quantum circuit optimization on a classical computer.
To that end, we propose and analyze specific classical algorithms to approximate the time evolution operator by shallow quantum circuits.
A previous proposal for this task focused on translationally invariant quantum systems and circuits~\cite{MaEtAl21}; we design our approach to be applicable to more general Hamiltonians and quantum circuits.
We make use of tensor network techniques~\cite{VeMuCi08, Or14, Ba22}.
More specifically, we write the quantum circuits needed for the optimization in terms of products of matrix product operators (MPOs)~\cite{VeGaCi04, ZwoVi04} and evaluate them by contracting the corresponding tensor networks.
For the gate optimization, we compare the coordinatewise approach in~\cite{BeFiLu21} to the Newton method.

\begin{figure}
\centering
\includegraphics[width=0.95\columnwidth]{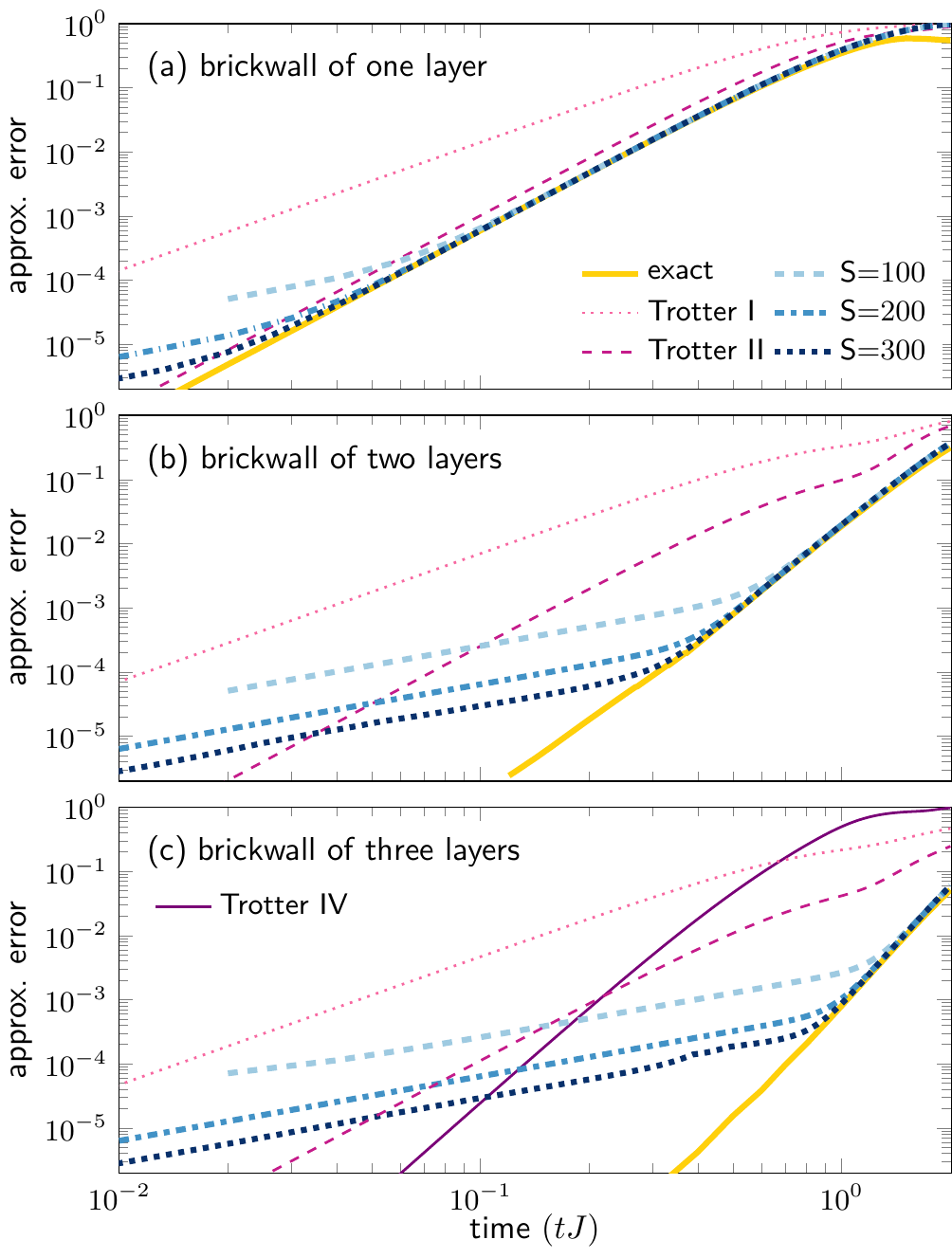}
\caption{\label{fig:1}
Classically optimized Hamiltonian simulation (thick lines) vs.\ Trotter product formulas (thin lines) for a brickwall circuit of depth (a) one, (b) two, and (c) three.
The approximation error, defined in Sec.~\ref{sec:Methods}, of the approximate evolution operator $U_{\text{approx}}$ with respect to the exact one $\exp(-\text{i} t H)$ is shown as a function of time $t$.
We consider $H = 2 \sum_{k = 1}^{7} Z_{k} Z_{k+1} + \sum_{k = 1}^{8} X_{k} + \sum_{k = 1}^{8} Z_{k}$.
For the Trotter results $U_{\text{approx}}$ is a Trotter product formula of $1$st (dotted), $2$nd (dashed) and $4$th (solid) order.
For the classically optimized results $U_{\text{approx}} = U(\boldsymbol{\theta})$ is the circuit with parameters $\boldsymbol{\theta}$ after two optimization procedures:
The first procedure minimizes the approximation error to the exact operator $\exp(-\text{i} t H)$ (solid); the second procedure cuts the total time into $S = 100$ (dashed), $200$ (dash-dotted) and $300$ (dotted) slices of equal time $\tau = 1 / S$ and then optimizes using a $1$st order Taylor approximation of $\exp(-\text{i} \tau H)$ and $S$ iterations thereof.
We observe that by increasing $S$ the results of the Taylor approach converge to the ones obtained via $\exp(-\text{i} t H)$.
Additionally we see that the classically optimized two- and three-layer circuits are two orders of magnitude more accurate than the Trotter formulas.
Further details are in Sec.~\ref{sec:Results}.
}
\end{figure}

These techniques enable us to efficiently work with various shallow ansatz circuits, e.g.\ of brickwall and sequential structure~\cite{LiEtAl20}, as well as with a large class of Hamiltonians, e.g.\ with open and periodic boundary conditions and with local and nonlocal interactions.
Additionally, the algorithms are hardware-agnostic and can handle any set of native quantum gates.
To illustrate the approach, we choose a brickwall circuit for the ansatz and an open quantum Ising chain with both transverse and longitudinal fields for the Hamiltonian.
The performance of some of our algorithms is shown in Fig.~\ref{fig:1}.

\clearpage

We see in Fig.~\ref{fig:1} that the Hamiltonian simulation for short times, i.e.\ $\exp(-\text{i} t H)$ for small values of $t$, is accurately represented by shallow circuits which can be efficiently run on classical computers.
To realize Hamiltonian simulation for longer times $T$, i.e.\ $\exp(-\text{i} T H)$ for a large value of $T$, we string together $T/t$ copies of the shallow circuit optimized for small $t$ and run them on a quantum computer, since $\exp(-\text{i} t H)^{T/t} = \exp(-\text{i} T H)$.

This procedure can be seen as an unconventional variant of hybrid quantum-classical algorithms.
Traditionally in hybrid quantum-classical approaches the quantum computer is used many times during the circuit optimization to evaluate the cost function and other quantities required for the optimization~\cite{McClean2016}.
We emphasize that in the traditional version of hybrid quantum-classical methods the quantum part can dominate the computational cost.
In our unconventional version, however, the quantum computer is used only once after the circuit optimization to run the circuits that were optimized entirely on a classical computer.

The article is structured as follows.
Section~\ref{sec:Methods} presents the algorithms and we analyze the results of our numerical experiments in Sec.~\ref{sec:Results}.
A concluding discussion is given in Sec.~\ref{sec:Discussion}.
The appendices provide additional information.

\section{Methods}
\label{sec:Methods}

To approximate the time evolution operator $\exp(-\text{i} t H)$ by a quantum circuit, we split the total evolution time $t$ into $S$ slices of equal size $\tau = t/S$ and then compute the circuit $U(\boldsymbol{\theta})$ for slice $s$ by minimizing the cost function $||U(\boldsymbol{\theta}) - \exp(-\text{i} \tau H) U_{s-1}||_{\text{F}}^{2}$ over the variational parameters $\boldsymbol{\theta} = (\theta_{1}, \theta_{2}, \ldots, \theta_{K})$.
Here $||\cdot||_{\text{F}}$ is the Frobenius norm, $U_{s-1}$ the final circuit after optimization of the previous slice and $U_{0} = \mathds{1}$.
The minimization of this cost function is equivalent to the maximization of the objective function
\begin{equation}\label{eq:Fidelity}
 F(\boldsymbol{\theta}) = \Re \{ \tr[ U^{\dag}(\boldsymbol{\theta}) \exp(-\text{i} \tau H) U_{s-1}] \}
\end{equation}
where $\Re \{ \cdot \}$ denotes the real part, $\tr[\cdot]$ the trace and $U^{\dag}(\boldsymbol{\theta})$ is the adjoint of $U(\boldsymbol{\theta})$.
We approximate $\exp(-\text{i} \tau H)$ by a truncated Taylor series and in our simulations use the MPO representation thereof~\cite{ZaEtAl15}.
This allows us to express the objective function as a product of MPOs, see App.~\ref{app:A} for details.
To quantify the circuit performance, we define the approximation error
\begin{equation}\label{eq:ApproxError}
 \epsilon_\text{approx} = \sqrt{1 - \Re \{ \tr[ U^{\dag}(\boldsymbol{\theta}) \exp(-\text{i} t H) ] \} / 2^{n}}
\end{equation}
for $n$ qubits which satisfies $\epsilon_\text{approx} \in [0, \sqrt{2}]$.
We also quantify the error in terms of the spectral norm in App.~\ref{app:B}.

We note that the quantities of interest, i.e.\ the approximation error~\eqref{eq:ApproxError} and the spectral norm error in App.~\ref{app:B}, are defined in terms of the exact time evolution operator $\exp(-\text{i} t H)$ and therefore their evaluation is exponentially costly in $n$.
The approximation error results presented here are therefore limited to rather small values of $n \leq 14$.

For circuits of qubit number greater than $n=14$, we quantify the circuit performance in terms of the (ground state) phase error \cite{VanDamme2023}
\begin{equation}\label{eq:PhaseError}
 \epsilon_\text{phase} = \lvert \lambda + \text{i} e_{0} t \rvert
\end{equation}
where $\lambda = \log \left( \bra{\psi_{0}} U(\boldsymbol{\theta}) \ket{\psi_{0}} \right)$ and $\ket{\psi_{0}}$ is the ground state of the problem Hamiltonian with energy $e_{0}$.
We evaluate $\ket{\psi_{0}}$ and $e_{0}$ using DMRG~\cite{RevModPhys.77.259} which gives numerically exact results for the Hamiltonian used.

The optimization has two key components: tensor network contraction and gate optimization.
Because we are using standard approaches for the tensor network contraction (as explained in App.~\ref{app:A}), in the following we focus on the gate optimization.

We propose to use a Newton method to maximize the objective function in Eq.~\eqref{eq:Fidelity}.
This method requires the gradient vector $\boldsymbol{\mathcal{G}}$ with elements $\mathcal{G}_{k} = \partial F / \partial \theta_{k}$ and Hessian matrix $\mathcal{H}$ with elements $\mathcal{H}_{j, k} = \partial^{2} F / \partial \theta_{j} \partial \theta_{k}$ and iterates
\begin{equation}\label{eq:Newton}
 \boldsymbol{\theta}^{(i+1)} = \boldsymbol{\theta}^{(i)} - \left( \mathcal{H}^{(i)} \right)^{-1} \boldsymbol{\mathcal{G}}^{(i)} .
\end{equation}
For the computation of the inverse of the Hessian matrix, we use that this matrix is Hermitian and that we are only interested in its negative eigenvalues, since our goal is the maximization of the objective function.
Therefore we compute the pseudoinverse via the eigendecomposition of the Hessian matrix and by setting all eigenvalues larger than some cutoff $-\epsilon$ to zero.

As an alternative to the Newton method, we also consider a coordinatewise optimization method which for the type of objective function Eq.~\eqref{eq:Fidelity} was derived in~\cite{BeFiLu21}.
The coordinatewise procedure updates variational parameters one after another and computes for each parameter:
\begin{eqnarray}\label{eq:Coordinatewise}
 \theta_{k}^{\text{new}} & = & \theta_{k}^{\text{old}} - 2\text{atan2}(F(\theta_{k}^{\text{old}}), F(\theta_{k}^{\text{old}}+\pi)) \nonumber\\
 & & + 4 \pi (p + \frac{1}{4}) \quad \forall p \in \mathbb{Z}\\
 F(\theta_{k}^{\text{new}}) & = & \sqrt{(F(\theta_{k}^{\text{old}}))^{2}+(F(\theta_{k}^{\text{old}}+\pi))^{2}}.
\end{eqnarray}
Here $\text{atan2}(y, x)$ returns the argument of the complex number $x + \text{i} y$.
We use the simplified notation \mbox{$F(\theta_{k}) = F(\theta_{1}, \ldots, \theta_{k-1}, \theta_{k}, \theta_{k+1}, \ldots, \theta_{K}) = F(\boldsymbol{\theta})$} for the objective function as a function of one specific parameter $\theta_{k}$ under the assumption that all other parameters $\theta_{j}$ for $j \neq k$ are fixed.
Note that $F(\theta_{k}^{\text{new}}) = F(\theta_{k+1}^{\text{old}})$.
Therefore, after the first parameter update (for $k = 1$) we only need one evaluation of the objective function at $\theta_{k}^{\text{old}}+\pi$ (for $k > 1$) for each parameter update, because we know $F(\theta_{k}^{\text{old}}) = F(\theta_{k-1}^{\text{new}})$ from the previous parameter update.
Note that the coordinatewise optimization is more efficient than the Newton method but expected to suffer more from getting stuck in local optima.

\section{Results}
\label{sec:Results}

\begin{figure}
\centering
\includegraphics[width=57.138mm]{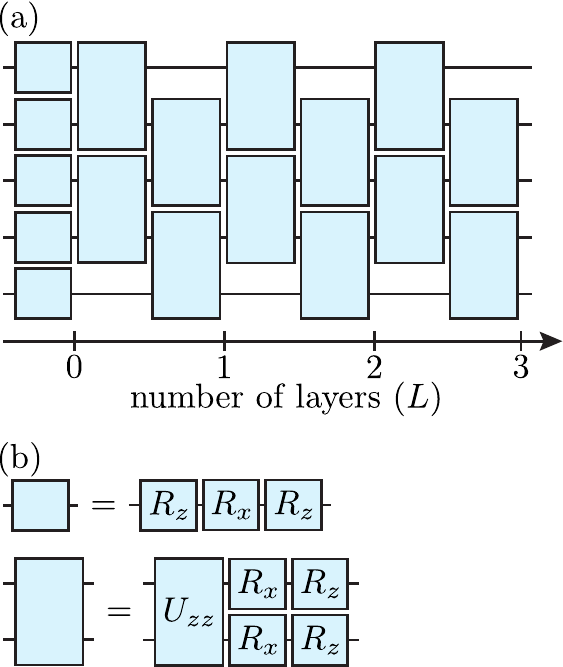}
\caption{\label{fig:2}
(a) Brickwall ansatz for $n = 5$ qubits.
(b) Parameterization in terms of the gates $R_{x}(\theta) = \exp(-\text{i} \theta X / 2)$, $R_{z}(\theta) = \exp(-\text{i} \theta Z / 2)$ and $U_{z z}(\theta) = \exp(-\text{i} \theta Z \otimes Z / 2)$ where $\theta$ denotes the variational parameter.
}
\end{figure}

We use the following setup in our numerical experiments.
For the Hamiltonian we consider the transverse-field quantum Ising chain with an additional magnetic field and open boundary conditions:
\begin{equation}\label{eq:H}
 H = J \sum_{k = 1}^{n-1} Z_{k} Z_{k+1} + g \sum_{k = 1}^{n} X_{k} + h \sum_{k = 1}^{n} Z_{k}
\end{equation}
where $X$ ($Z$) is the Pauli $X$ ($Z$) matrix and the Hamiltonian parameters $J$, $g$ and $h$ are set to the non-integrable point $(J,g,h) = (2.0,1.0,1.0)$ throughout the article.
For the quantum circuit ansatz we use the brickwall structure shown in Fig.~\ref{fig:2} (a) composed of the gates shown in Fig.~\ref{fig:2} (b).
Newton's method is used with an eigenvalue cutoff of $10^{-5}$ and iterated until the Euclidean norm of the vector of gradients falls below $10^{-5}$.
The tensor network calculations make use of the libraries~\cite{TensorKit,pastaq}.

\begin{figure}
\centering
\includegraphics[width=0.95\columnwidth]{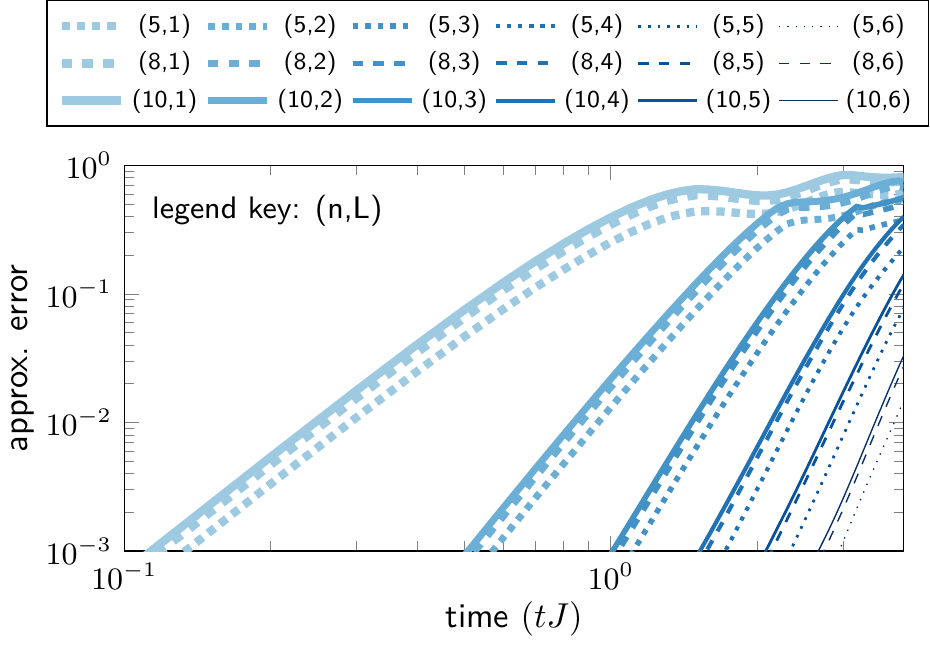}
\caption{\label{fig:3}
Approximation error as a function of time for classically optimized brickwall circuits of $L$ layers and $n$ qubits.
}
\end{figure}

\begin{figure}
\centering
\includegraphics[width=0.95\columnwidth]{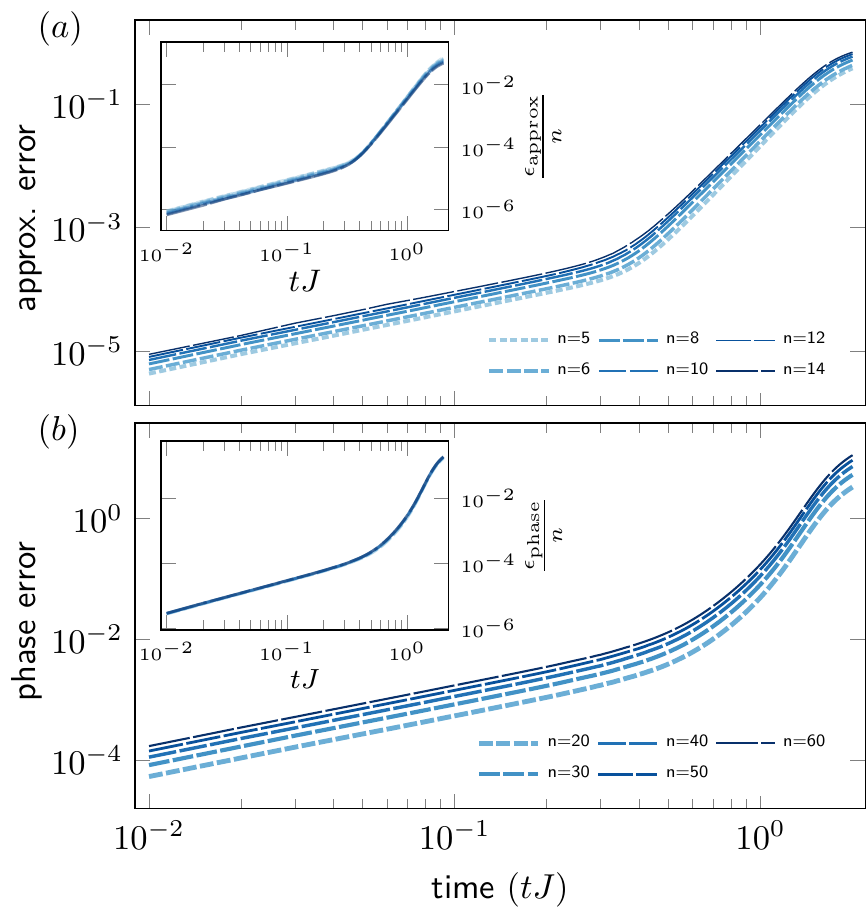}
\caption{\label{fig:4}
(a) Approximation error and (b) phase error of the classically optimized circuit $U(\boldsymbol{\theta})$ achieved for various qubit counts $n$.
We cut the total time into $S = 200$ slices of equal time $\tau = 1 / S$ and then optimize using L-BFGS, a $1$st order Taylor approximation of $\exp(-\text{i} \tau H)$ and $S$ iterations thereof.
We consider the Hamiltonian in Eq.~\eqref{eq:H} and a brickwall circuit of two layers $(L=2)$.
The insets show the approximation (a) and phase (b) errors normalized by the number of qubits $n$.
In both cases, the collapse of the data onto a single curve indicates a linear scaling of the errors with qubit number.
}
\end{figure}

We compare various gate optimization methods including global and batched Newton methods and the coordinatewise scheme in App.~\ref{app:C}.
Because the global Newton method performs best, we use this method in the classical optimization of small systems; for example, the results in Fig.~\ref{fig:1}.
For larger circuits with more parameters we find that the quasi-Newton, Broyden-Fletcher-Goldfarb-Shanno (BFGS) algorithm~\cite{Nocedal2009} and its limited-memory counterpart L-BFGS~\cite{Liu1989} produce satisfactory results without requiring explicit calculation of the Hessian.
We therefore use L-BFGS to calculate the results in Figs.~\ref{fig:3} and~\ref{fig:4}.

We assess the performance of the classical algorithms by analyzing the results shown in Fig.~\ref{fig:1}.
Let us first explain what is shown in Fig.~\ref{fig:1}.
For the Trotter product formulas, we define $H_{Z} = J \sum_{k = 1}^{n-1} Z_{k} Z_{k+1} + h \sum_{k = 1}^{n} Z_{k}$ and $H_{X} = g \sum_{k = 1}^{n} X_{k}$ and use the $1$st order (\textsf{Trotter I}), $2$nd order (\textsf{Trotter II}) and $4$th order (\textsf{Trotter IV}) formulas
\begin{eqnarray*}
e^{-\text{i} t H_{X}} e^{-\text{i} t H_{Z}} & \quad & (\textsf{Trotter I}),\\
e^{-\text{i} t H_{X} / 2} e^{-\text{i} t H_{Z}} e^{-\text{i} t H_{X} / 2} & \quad & (\textsf{Trotter II}),\\
e^{-\text{i} s t H_{X} / 2} e^{-\text{i} s t H_{Z}} e^{-\text{i} (1-s) t H_{X} / 2} e^{-\text{i} (1-2s) t H_{Z}}\\
e^{-\text{i} (1-s) t H_{X} / 2} e^{-\text{i} s t H_{Z}} e^{-\text{i} s t H_{X} / 2} & \quad & (\textsf{Trotter IV})
\end{eqnarray*}
where $s = 1/(2-\sqrt[3]{2})$~\cite{HaSu05}.
Note that $e^{-\text{i} t H_{Z}} = \prod_{k = 1}^{n-1} e^{-\text{i} t J Z_{k} Z_{k+1}} \prod_{k = 1}^{n} e^{-\text{i} t h Z_{k}}$ and $e^{-\text{i} t H_{X}} = \prod_{k = 1}^{n} e^{-\text{i} t g X_{k}}$ are products of exponentials that have an exact representation using the gates of our ansatz in Fig.~\ref{fig:2}.
Also note that only the exponentials in $\prod_{k = 1}^{n-1} e^{-\text{i} t J Z_{k} Z_{k+1}}$ are two-qubit gates and all the other exponentials are one-qubit gates.
For a brickwall ansatz of one layer, \textsf{Trotter I} and \textsf{Trotter II} are used for the Trotter results.
The same is true for a brickwall ansatz of two layers, whereby two identical Trotter layers, each corresponding to an evolution by a time $t/2$, specify the circuit.
Similarly, for a brickwall ansatz of three layers, three identical Trotter layers are composed to specify the circuit.
Furthermore, a circuit of three layers allows us to also use \textsf{Trotter IV} for the Trotter product formula.
The approximation errors of the Trotter based circuits are plotted in Fig.~\ref{fig:1} and facilitate a fair comparison with the classically optimized circuits inasmuch as the depth of two-qubit gates in the circuits considered within each subfigure -- Fig.~\ref{fig:1} (a), (b) and (c) -- are the same.

In order to estimate how well the optimization scheme can exhaust the potential of the brickwall circuits, we first optimize the variational circuit parameters for a cost function which is constructed using the exact unitary $\exp(-\text{i} t H)$ and is calculated via matrix exponentiation.
Supposing that the Newton based optimization scheme finds the global maximum of the objective function, these results (denoted \textsf{exact} in Fig.~\ref{fig:1}) give the minimum approximation error which can be achieved using the respective circuits.
Compared to the Trotter based approximation schemes, the improvement using the \textsf{exact} scheme is small for a brickwall circuit of one layer (Fig.~\ref{fig:1} a), however, deeper circuits (Fig.~\ref{fig:1} b and c) lead to approximation errors which are orders of magnitude smaller than those obtained using Trotter product formulas.

The task of the Taylor based approach is to find circuit approximations as close as possible to the optimum achieved using the \textsf{exact} approach, in a manner which does not require exponentiation of the full Hamiltonian and is therefore scalable to larger system sizes.
We find that the classical optimization scheme using a $1$st order Taylor approximation is capable of finding variational parameters which approximate the target unitary with an approximation error equivalent to the \textsf{exact} scheme.
This is evident in Fig.~\ref{fig:1} where the \textsf{exact} and sliced Taylor (\mbox{\textsf{S = 100}}, \textsf{S = 200}, \textsf{S = 300}) results overlap, indicating that the approximation scheme has saturated the expressiveness of the circuit ansatz.
Additionally, the time at which this saturation occurs falls earlier if more slices $S$ are used.
We have also performed all the above calculations for a system of $n = 5$ qubits, see App.~\ref{app:D}, and the results are quantitatively similar to the $n = 8$ results presented here.

The slopes of the data presented in Fig.~\ref{fig:1} can be understood by considering the leading error arising from each approximating scheme.
The leading errors of the Trotter product formulas are known to be of orders $O(t^{2})$, $O(t^{3})$ and $O(t^{5})$ for the $1$st, $2$nd and $4$th order Trotter product formulas, respectively, and this is apparent in the slope of the corresponding data presented in Fig.~\ref{fig:1}.
For small times, the data for \textsf{Trotter I}, \textsf{Trotter II} and \textsf{Trotter IV} have slopes approximately equal to $m = 2$ , $m = 3$ and $m = 5$, respectively.

The approximation errors arising from circuits which have been classically optimized via Taylor (\textsf{S = 100}, \mbox{\textsf{S = 200}}, \textsf{S = 300}) are based on a $1$st order Taylor approximation.
These curves are split into two regimes.
The first regime occurs before the data overlap with those of the \textsf{exact} method and have a slope approximately equal to $m = 1$, indicating a leading error of order $O(t)$, as expected for the error accumulated by a $1$st order Taylor approximation.
The second regime occurs at larger times where the Taylor data overlap with the \textsf{exact} data.
Here the corresponding slopes reflect the expressiveness of the circuits used.

To examine how the errors scale due to the expressiveness of the ansatz, we calculate the approximation error of circuits optimized using the \textsf{exact} method for several values of $n$ and $L$; the results are plotted in Fig.~\ref{fig:3}.
We estimate the error scalings by fitting $\epsilon_\text{approx}(t) = c(n) t^{m}$ to the data as explained in App.~\ref{app:E}.
The resulting prefactors $c$ and exponents $m$ for each circuit size $(n,L)$ are provided in Tab.~\ref{tab:1} alongside those of circuits constructed using Trotter product formulas.
We find that the classically optimized circuits can significantly outperform Trotter product formulas in both the prefactors and the exponents.
Additionally we see in Fig.~\ref{fig:3} that, for a fixed approximation error, the maximum achievable time appears to scale linearly with $L$ which amounts to the best scaling possible~\cite{Berry2007a}.

\begin{table}
\small
\renewcommand{\arraystretch}{1.2}
\caption{\label{tab:1}
Comparison of error scalings $\epsilon_\text{approx} = c(n)t^{m}$ for circuits approximating the unitary $\exp(-\text{i} t H)$ constructed using classical optimization and Trotter product formulas~\cite{Suzuki1990, Suzuki1991}.
The prefactors $c(n)$ (lower table) and exponents $m$ (upper table) are found by fitting to the approximation error (see App.~\ref{app:E} for details).
Note that the $L = 5$ Trotter product formula is constructed using the recursion relation in Eq.~(3.14) of~\cite{Suzuki1991}.
}
\begin{ruledtabular}
\begin{tabular}{@{}lrrrrrr@{}}
\multicolumn{1}{@{}}{} &
\multicolumn{3}{c}{classically optimized} &
\multicolumn{3}{c}{Trotter}\\
\cmidrule(r{0.1em}){2-4}
\cmidrule(l{0.1em}){5-7}
L&
\multicolumn{1}{c}{n=5}&
\multicolumn{1}{c}{n=8}&
\multicolumn{1}{c}{n=10}&
\multicolumn{1}{c}{n=5}&
\multicolumn{1}{c}{n=8}&
\multicolumn{1}{c}{n=10}\\
\cmidrule(){2-2}
\cmidrule(){3-3}
\cmidrule(r{0.1em}){4-4}
\cmidrule(l{0.1em}){5-5}
\cmidrule(){6-6}
\cmidrule(){7-7}
1 & $2.9(1)$ & $2.9(1)$ & $2.9(1)$ & $1.98(3)$ & $1.98(4)$ & $1.98(4)$ \\
1 & -        & -        & -        & $2.98(4)$ & $2.98(4)$ & $2.98(3)$ \\
2 & $4.6(2)$ & $4.6(1)$ & $4.6(2)$ & -   & -   & -         \\ 
3 & $6.2(1)$ & $6.2(1)$ & $6.2(1)$ & $4.92(6)$ & $4.95(9)$ & $5.0(1)$  \\
4 & $7.4(1)$ & $7.0(3)$ & $7.0(1)$ & -   & -   & -         \\
5 & $8.3(2)$ & $7.9(3)$ & $7.8(3)$ & $4.98(2)$ & $4.99(3)$ & $4.98(3)$ \\
6 & $9.0(3)$ & $8.9(4)$ & $8.7(6)$ & -   & -   & -         \\
\cmidrule(l{0.1em}r{0.1em}){2-7}
1 & ${\scriptstyle3.5(3)\cdot10^{-1}}$ & ${\scriptstyle5.1(6)\cdot10^{-1}}$ & ${\scriptstyle5.9(7)\cdot10^{-1}}$ & ${\scriptstyle1.02(6)}$  & ${\scriptstyle1.33(8)}$ & ${\scriptstyle1.5(1)}$  \\
1 & -                                  & -                                  & -                                  & ${\scriptstyle0.71(4)}$  & ${\scriptstyle0.96(6)}$ & ${\scriptstyle1.10(7)}$ \\
2 & ${\scriptstyle1.3(1)\cdot10^{-2}}$ & ${\scriptstyle1.9(1)\cdot10^{-2}}$ & ${\scriptstyle2.2(1)\cdot10^{-3}}$ & -                        & -                       & -                       \\ 
3 & ${\scriptstyle5.2(3)\cdot10^{-4}}$ & ${\scriptstyle8.1(3)\cdot10^{-4}}$ & ${\scriptstyle9.5(1)\cdot10^{-4}}$ & ${\scriptstyle1.5(2)}$   & ${\scriptstyle2.2(4)}$  & ${\scriptstyle2.6(5)}$  \\
4 & ${\scriptstyle1.8(1)\cdot10^{-5}}$ & ${\scriptstyle4.1(6)\cdot10^{-5}}$ & ${\scriptstyle5.2(4)\cdot10^{-5}}$ & -                        & -                       & -                       \\
5 & ${\scriptstyle9(2)\cdot10^{-7}}$   & ${\scriptstyle2.5(6)\cdot10^{-6}}$ & ${\scriptstyle3.2(9)\cdot10^{-6}}$ & ${\scriptstyle0.107(3)}$ & ${\scriptstyle0.153(8)}$& ${\scriptstyle0.176(8)}$\\
6 & ${\scriptstyle6(2)\cdot10^{-8}}$   & ${\scriptstyle1.2(7)\cdot10^{-7}}$ & ${\scriptstyle2(2)\cdot10^{-7}}$   & -                        & -                       & -                       \\
\end{tabular}
\end{ruledtabular}
\end{table}

To illustrate the behavior of the Taylor based optimization with respect to system size, the data in Fig.~\ref{fig:4} compare the approximation errors achieved using the (\textsf{S = 200}) Taylor based method for systems of $n = 5$, $6$, $8$, $10$, $12$ and $14$ qubits.
We find that the approximation error has a very similar behavior for all system sizes and when they are normalized by the number of qubits, the data collapse onto a single curve (Fig.~\ref{fig:4} (a) inset) indicating a linear scaling of the approximation error with qubit number.
Furthermore, the turning point at which the slope changes falls at approximately the same time for all values of $n$.
Note that the turning point of the approximation error tells us where the classical optimization has the largest advantage over Trotter product formulas, cf.\ Fig.~\ref{fig:1}.
The computation of the approximation error, however, is based on the exact exponential $\exp(-\text{i} t H)$ and, therefore, is not efficient for large systems.
Figure~\ref{fig:4} suggests that data from small systems may be used to inform the classical optimization of larger circuits where calculation of the approximation error is no longer feasible.
In particular, data from small systems may be used to determine the number of slices which should be performed in order to reach the turning point.

To go beyond the limitations imposed by the calculation of the approximation error we also calculate the phase error defined in Eq.~(\ref{eq:PhaseError}) for systems of $n = 20$, $30$, $40$, $50$ and $60$ qubits for the same parameters used in Fig.~\ref{fig:4} (a) i.e.\ $S=200$, $L=2$.
The resulting data are shown in Fig.~\ref{fig:4} (b).
The phase error data are qualitatively similar to the approximation error data: at early times the phase error accumulates due to the Taylor approximation error whereas at larger times the error is related to the expressiveness of the ansatz.
Furthermore, the phase error data normalized by the number of qubits also collapse onto a single curve (Fig.~\ref{fig:4} (b) inset) indicating a linear scaling of the phase error with $n$.

Let us address the question how many gates are needed to achieve a desired total error $\epsilon_{\text{total}}$ for a total Hamiltonian simulation time $T$.
We know from our simulations that the approximation error has the form $\epsilon_{L}(t) = c_{L}(n) t^{m_{L}}$ where $c_{L}(n)$ and $m_{L}$ are positive real numbers that are determined by the number of brickwall layers $L$ in the ansatz.
In particular, we observe in Fig.~\ref{fig:3} that for small approximation errors, the error scales as a polynomial in time $t^{m_L}$ with an exponent which is independent of the qubit number $n$ while the prefactors $c_{L}(n)$ have a dependence on the number of qubits.
We realize Hamiltonian simulation for a total time $T$ by appending $r=T/t$ circuits each approximating Hamiltonian simulation for $t \ll T$ with error $\epsilon_{L}$.

The total error is that accumulated by repeating the circuit $r$ times
\begin{equation}\label{eq:TrotterNumberInequality}
\begin{split}
\epsilon_\text{total} &= \norm{U^r_{L}(T/r) - e^{-\text{i}TH}} \\&\leq r\norm{U_{L}(T/r) - e^{-\text{i}TH/r}}\\ &=rc_{L}(n)(T/r)^{m_{L}}
\end{split}
\end{equation}
where $U_{L}(T/r)$ is an $L$-layer classically optimized brickwall circuit which approximates the time evolution operator $\exp(-\text{i}TH/r)$.
We solve Eq.~\eqref{eq:TrotterNumberInequality} for the number of circuit repetitions $r$ which gives, for $m_{L} > 1$
\begin{equation}
 r \leq \left( \frac{c_{L}(n)}{\epsilon_{\text{total}}} \right)^{\frac{1}{m_{L}-1}} T^{\frac{m_{L}}{m_{L}-1}}.
\end{equation}

The total number of gates $N_{\text{gates}}$ based on the approximation of $\exp(-\text{i} t H)$ by $L$ brickwall layers is $N_{\text{gates}} = r \cdot (n-1)L$ (neglecting single-qubit gates) and using the previous result for $r$ becomes
\begin{eqnarray}\label{eq:NumberGates}
 N_{\text{gates}} & \leq & (n-1) L \left( \frac{c_{L}(n)}{\epsilon_{\text{total}}} \right)^{\frac{1}{m_{L} - 1}} T^{\frac{m_{L}}{m_{L} - 1}}.
\end{eqnarray}
The data presented in Fig.~\ref{fig:4} indicate that $c_{L}(n) \propto n$ giving a gate complexity $O((n T)^{1+\frac{1}{m_{L}-1}})$ which is consistent with known asymptotic scalings of product formulas for geometrically local Hamiltonians \cite{PhysRevLett.123.050503}.
Note that by solving Eq.~\eqref{eq:NumberGates} for $T$ we get
\begin{eqnarray}\label{eq:TotalSimulationTime}
 T & \geq & \left( \frac{N_{\text{gates}}}{(n-1) L} \right)^{\frac{m_{L} - 1}{m_{L}}} \left( \frac{\epsilon_{\text{total}}}{c_{L}(n)} \right)^{\frac{1}{m_{L}}},
\end{eqnarray}
i.e.\ the total simulation time feasible for given total error and gate count.

To compare $N_{\text{gates}}$ in Eq.~\eqref{eq:NumberGates} and $T$ in Eq.~\eqref{eq:TotalSimulationTime} of the classically optimized circuits with the corresponding values of Trotter product formulas, we note that the Trotter error has the same functional form as the classically optimized approximation error, i.e.\ $\epsilon_{L}^{\text{Trotter}}(t) = c_{L}(n)^{\text{Trotter}} t^{m_{L}^{\text{Trotter}}}$.
Therefore $N_{\text{gates}}$ for the Trotter product formulas has the same form as Eq.~\eqref{eq:NumberGates} and $T$ for the Trotter product formulas has the same form as Eq.~\eqref{eq:TotalSimulationTime}.
So we only need to compare the values of $c_{L}(n)$ and $m_{L}$ with $c_{L}(n)^{\text{Trotter}}$ and $m_{L}^{\text{Trotter}}$, respectively.
This comparison is done in Tab.~\ref{tab:1} where we see that $c_{L}(n) < c_{L}(n)^{\text{Trotter}}$ and $m_{L} > m_{L}^{\text{Trotter}}$ for $L > 1$.
We conclude that the classically optimized circuits can lead to a considerable reduction of $N_{\text{gates}}$ for given $\epsilon_{\text{total}}$ and $T$.
Additionally we conclude that they can lead to a significant increase of $T$ for given $N_{\text{gates}}$ and $\epsilon_{\text{total}}$.

\section{Discussion}
\label{sec:Discussion}

In this article we present classical algorithms that reduce the quantum hardware requirements for simulating the time evolution of quantum systems and we illustrate their performance for a local Hamiltonian with open boundary conditions using a variational brickwall quantum circuit.
These algorithms can readily tackle a wider range of problems and optimize other types of quantum circuits.
For example, periodic systems, Hamiltonians with nonlocal interactions or the sequential quantum circuit~\cite{LiEtAl20} can be considered.
Even shallow two-dimensional quantum circuits, as required by some of the IBM quantum devices, can be handled by making use of tensor network techniques for two-dimensional systems, e.g.~\cite{VeCi04, LuCiBa14a, LuCiBa14b, CzDzCo19, McSz21}.

To efficiently work with time evolution over long times, we cut the total time evolution into several slices and then represent the shorter time evolution per slice using a $1$st order Taylor approximation.
It is interesting to study more accurate approximations~\cite{Os22}, e.g.\ a higher-order truncated Taylor series or perhaps a Jacobi-Anger expansion~\cite{AbSt66}, that decrease the number of slices needed for a certain accuracy.
Any approximation of the time evolution operator as a finite polynomial $\exp(-\text{i} t H) \approx \sum_{k} a_{k} H^{k}$ leads to an objective function that is a finite sum $F(\boldsymbol{\theta}) = \Re \{ \tr[ U^{\dag}(\boldsymbol{\theta}) \exp(-\text{i} t H) ] \} \approx \sum_{k} a_{k} \Re \{ \tr[ U^{\dag}(\boldsymbol{\theta}) H^{k} ] \}$.
The question is whether the additional cost of evaluating higher powers of $H$ can still lead to a reduction of the overall cost due to the smaller total number of slices.

We consider a variety of methods in this article for the numerical optimization of circuit parameters, ranging from a simple coordinatewise update scheme to a more sophisticated global Newton method.
The global Newton approach outperformed all the other optimization procedures but is characterized by the largest computational cost per update due to the computation and inversion of the Hessian matrix.
We have also analyzed the performance of the more efficient quasi-Newton methods BFGS~\cite{Nocedal2009} and L-BFGS~\cite{Liu1989} which do not need the explicit calculation and inversion of the Hessian matrix.
We find that they both perform well and, although their convergence with the number of updates is slower, their advantage over the Newton method becomes apparent as the system size grows.
Additionally combining the (quasi-)Newton approaches with automatic differentiation~\cite{LiEtAl19} leads to a particularly efficient approach.

Finally, let us also address the question of how do gate errors in current quantum devices affect our results.
To model the gate errors, let us apply the two-qubit depolarizing channel~\cite{NiCh10} after each two-qubit gate so that each two-qubit gate has an error probability $p$.
Then, for $K$ two-qubit gates and sufficiently small $p \ll 1$, the total gate error is approximately $1-(1-p)^{K} \approx p K$.
To be more specific, we set $K = (n-1)L$ according to Fig.~\ref{fig:2} and consider the current Quantinuum quantum computer where $p \approx 10^{-3}$~\cite{RyEtAl22}.
For the problems considered in this article, therefore, the total gate errors are in between $\approx 10^{-3}$ and $\approx 10^{-1}$.
It is important to emphasize that the gate errors lower-bound the achievable accuracy when the circuits run on real quantum hardware.
This implies that, when the comparison is done on actual quantum computers, the advantage of the classically optimized circuits over Trotter product formulas depends on the gate error probability $p$ and increases for decreasing values of $p$, which is discussed in more detail in App.~\ref{app:F}.
Because the total gate errors are relatively large compared with the approximation errors, we can further speed up the classical optimization methods by using approximations such as approximate tensor network contractions and relaxed convergence conditions.

After completion of this work we learned about the related study~\cite{TeHaLu22}.

\section{Acknowledgments}

We are thankful for valuable discussions with David Amaro, Marcello Benedetti, Yuta Kikuchi and Chris N.\ Self.

\appendix

\section{Tensor network algorithms}
\label{app:A}

In this appendix we provide further details about the tensor network methods used to efficiently evaluate and optimize the cost function of Eq.~\eqref{eq:Fidelity}.

\begin{figure}
\centering
\includegraphics[width=77.575mm]{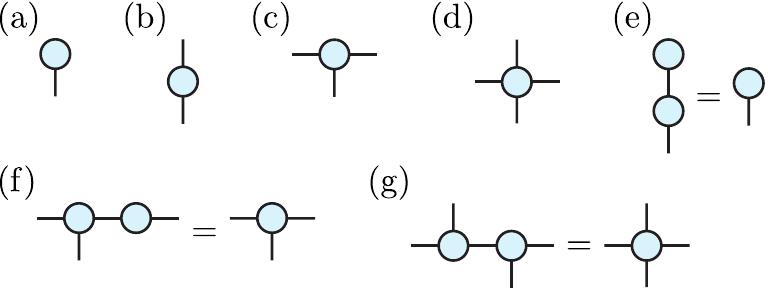}
\caption{\label{fig:5}
Basic examples of tensor network diagrams.
(a) A vector corresponds to a node with one leg which represents the index of the vector.
(b) A matrix has two legs for the row and column indices of the matrix.
(c) A third order tensor has three legs for three indices.
(d) A fourth order tensor has four legs for four indices.
(e) The product of a matrix and a vector gives another vector.
The summation in the matrix-vector product over the connecting indices between the matrix and the vector is represented by the connecting edge between the nodes.
(f) The tensor contraction of a third order tensor with a matrix over one index gives a third order tensor.
(g) The tensor contraction of two third order tensors over one index gives a fourth order tensor.
}
\end{figure}

To describe our approach it is convenient to use the diagrammatic notation to represent tensor networks.
Figure~\ref{fig:5} shows examples of that notation, see~\cite{VeMuCi08, Or14, Ba22} for additional details.

\begin{figure}
\centering
\includegraphics[width=77.725mm]{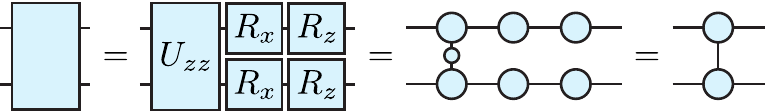}
\caption{\label{fig:6}
The two-qubit gate of our ansatz is composed of the two-qubit gate $U_{zz}$ and the one-qubit gates $R_x$ and $R_z$.
The one-qubit gates are represented as matrices in the tensor notation while the two-qubit $U_{zz}$ gate is decomposed into a pair of third order tensors and a bond matrix which connects them according to Eq.~\eqref{eq:SUzz}.
These tensors are contracted to give a pair of third order tensors which represents the entire two-qubit gate of our ansatz.
}
\end{figure}

\begin{figure}
\centering
\includegraphics[width=70.101mm]{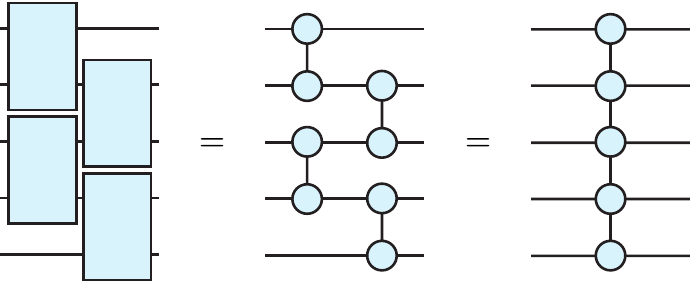}
\caption{\label{fig:7}
Decomposition of one brickwall layer of our ansatz into a MPO for $n = 5$ qubits.
Each two-qubit gate is first decomposed into a pair or third order tensors as described in Fig.~\ref{fig:6}.
Then we contract adjacent tensors via the horizontally connecting edges to form a MPO.
}
\end{figure}

\begin{figure}
\centering
\includegraphics[width=70.021mm]{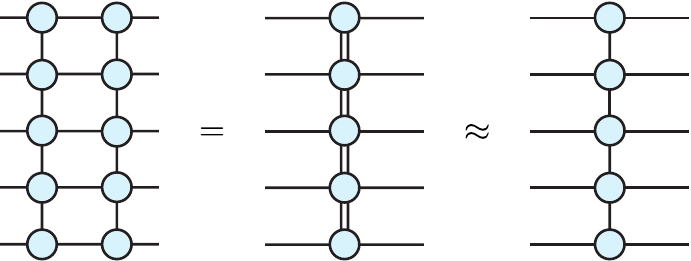}
\caption{\label{fig:8}
A pair of MPOs is represented by a single new MPO.
We contract the MPOs such that the bond dimension of the new MPO is the product of those of the contracted MPOs as shown in the central figure.
We find an approximate MPO representation of smaller bond dimension using the compression techniques in~\cite{VeGaCi04}.
}
\end{figure}

Since the cost function Eq.~\eqref{eq:Fidelity} contains the circuit ansatz $U(\boldsymbol{\theta})$ it is useful to express this circuit as a tensor network.
To do this, we start at the level of the individual two-qubit gates of our ansatz shown in Fig.~\ref{fig:6}.
We decompose the two-qubit gate
\begin{eqnarray}\label{eq:SUzz}
U_{z z}(\theta) & = & \exp(-\text{i} \theta Z \otimes Z / 2)\\\nonumber
& = & \cos(\theta/2) \mathds{1} \otimes \mathds{1} - \text{i} \sin(\theta/2) Z \otimes Z
\end{eqnarray}
where $\mathds{1}$ is the one-qubit identity operator.
This decomposition has a tensor network representation in terms of two third order tensors and one matrix as shown in Fig.~\ref{fig:6}.
We multiply this matrix with one of the third order tensors.
Then we multiply the $R_{x}$ and $R_{z}$ matrices with their adjacent third order tensors.
The result is the two-qubit matrix product operator (MPO)~\cite{VeGaCi04, ZwoVi04} of bond dimension $\chi = 2$ depicted as the right-most tensor network in Fig.~\ref{fig:6}.
Using this two-qubit gate decomposition, a single layer of our brickwall ansatz can be represented in terms of a MPO of bond dimension $\chi = 2$ as shown in Fig.~\ref{fig:7}.
Figure~\ref{fig:8} shows that we can exactly represent two layers of our ansatz by a single MPO of bond dimension $\chi = 4$ or approximately by a MPO of smaller bond dimension $\chi < 4$ using tensor network compression techniques~\cite{VeGaCi04}.
For $L$ layers of our ansatz, by repeating the procedure in Fig.~\ref{fig:8} $L$ times, we obtain a MPO representation that is exact if $\chi = 2^{L}$ or approximate if $\chi < 2^{L}$.

To approximate the unitary operator $\exp(-\text{i} \tau H)$ as an MPO for some small time step $\tau = t / S$ we use techniques from~\cite{ZaEtAl15}.
There the authors show how to construct the so-called $W^{I}$ and $W^{II}$ compact MPO approximations to the unitary $\exp(-\text{i} t H)$ directly using the MPO representation of the Hamiltonian $H$.
For a system of size $n$ both the $W^{I}$ and $W^{II}$ approximations have a total error which scales formally as $O(n \tau^{2})$ and a maximum MPO bond dimension $\tilde{\chi} = D-1$ where $D$ is the maximum bond dimension of the MPO representation of the Hamiltonian.

\begin{figure}
\centering
\includegraphics[width=58.032mm]{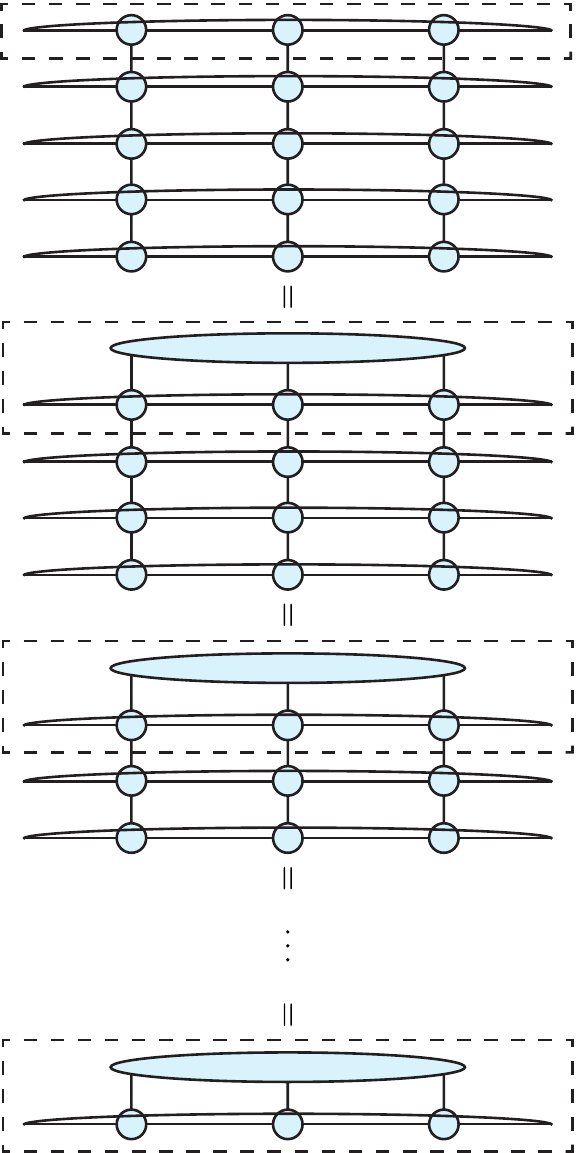}
\caption{\label{fig:9}
Trace over a product of three MPOs as required for the evaluation of Eq.~\eqref{eq:Fidelity}.
We first contract the three tensors in the top-most dashed rectangle into one large tensor represented by an oval node with three indices.
Then we contract this large tensor with the three adjacent tensors one after another.
The procedure is repeated $O(n)$ times until the final contraction with the bottom-most three tensors gives the desired scalar number representing the trace over the MPO product.
}
\end{figure}

Let us now discuss the tensor network for Eq.~\eqref{eq:Fidelity} and the computational cost of evaluating it.
We have seen above that $U(\boldsymbol{\theta})$ is a MPO of bond dimension $\chi \leq 2^{L}$ and $\exp(-\text{i} \tau H)$ is a MPO of bond dimension $\tilde{\chi}$.
The circuit $U_{s-1}$ is the optimized ansatz from the previous step which we represent by another MPO of bond dimension $\chi \leq 2^{L}$ equivalent to that of $U(\boldsymbol{\theta})$.
We exactly contract the trace of the product of the three MPOs for $U(\boldsymbol{\theta})$, $\exp(-\text{i} \tau H)$ and $U_{s-1}$ following the steps illustrated in Fig.~\ref{fig:9} whereby the network is contracted sequentially from top to bottom.
It is straightforward to see that, by performing the contraction one tensor at a time, the computational cost of this procedure scales as $O(n \tilde{\chi}^{2} \chi^{2}) + O(n \tilde{\chi} \chi^{3})$ which is linear in $n$ and where $\chi \leq 2^{L}$.
We note that for the Hamiltonian of Eq.~\eqref{eq:H} $\tilde{\chi} = 2$.
In all of our optimizations we choose $\chi \leq 2^{L}$ large enough such that any approximations are negligible.

Gradients of the objective function are computed using the same tensor network contraction by first replacing the $j$th variational Pauli rotation $R_j(\theta_j) = \exp(-\text{i} \theta P_{j} / 2)$ by its gradient, i.e.\ for $\partial F / \partial \theta_{j}$ replace $R_j(\theta_j)$ with $\partial R_j / \partial \theta_{j} = -\text{i} P_j R_j(\theta_{j}) / 2$.
A similar procedure can be used to calculate the Hessian $\partial^{2} F / \partial \theta_{j} \partial \theta_{k}$.
For a circuit with $Q$ parameters, this naive procedure incurs a computational cost which is $Q$ times the cost of contracting the network for $F$ as described in the previous paragraph; similarly for the independent terms in the Hessian matrix.
However, the efficiency of calculating gradients and the Hessian can be improved considerably by computing their entries in parallel or by caching and reusing tensor environments rather than performing the full tensor network contraction for each gradient or Hessian term.
Automatic differentiation can also be used to calculate derivatives and may simplify the task significantly~\cite{LiEtAl19}.

\section{Spectral norm}
\label{app:B}

As an alternative to the approximation error of Eq.~\eqref{eq:ApproxError}, in this appendix we study the spectral-norm distance
\begin{equation}\label{eq:SSpecDist}
 \epsilon_{\text{spec}} = \|\exp(-\text{i} t H) - U_\text{approx}\|
\end{equation}
where $\|(\cdot)\|$ denotes the spectral norm of $(\cdot)$.

\begin{figure}
\centering
\includegraphics[width=0.95\columnwidth]{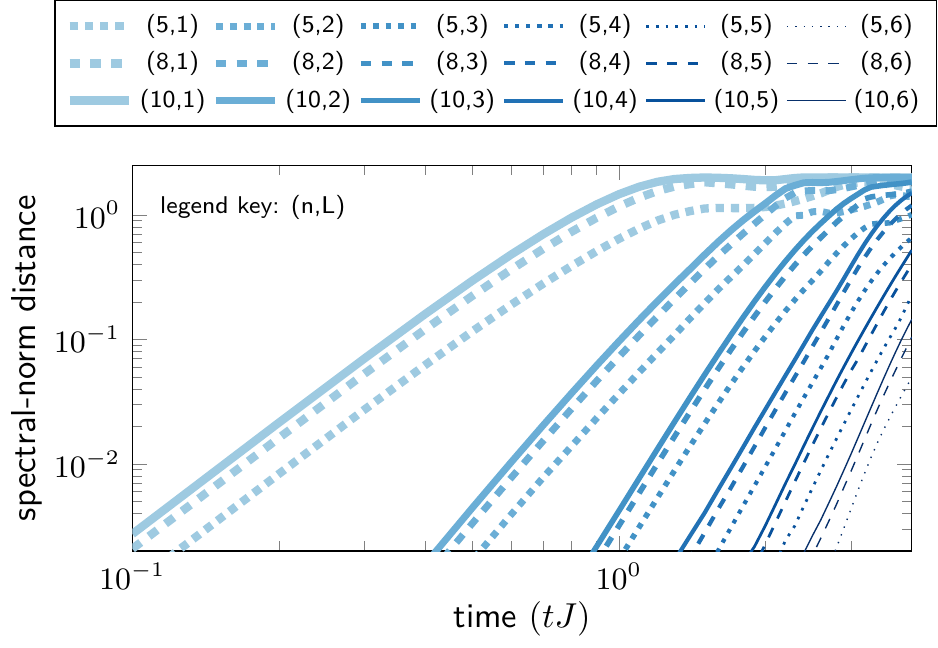}
\caption{\label{fig:10}
Spectral-norm distance as a function of time for classically optimized brickwall circuits of $L$ layers, $n$ qubits and for the Hamiltonian of Eq.~\eqref{eq:H} with parameters $(J,g,h) = (2.0,1.0,1.0)$.
The classical optimization was performed using the \textsf{exact} method outlined in Sec.~\ref{sec:Results}.
A constant factor increase in the spectral-norm distance is observed as $n$ increases while the error scalings (slopes of the linear parts of the curves) are approximately independent of the system size.
}
\end{figure}

Figure~\ref{fig:10} shows the spectral-norm distance as a function of time for the same set of circuits $n \in [5, 8, 10]$ and $L \in [1, 2, 3, 4, 5, 6]$ and for the same Hamiltonian considered in Fig.~\ref{fig:3}.
By fitting $\epsilon_{\text{spec}} = c t^{m}$ to the data, we get the prefactors and exponents given in Tab.~\ref{tab:2}.
The exponents of the spectral-norm distance curves are approximately equal to those of the approximation error whereas the prefactors tend to be larger, cf.\ Tab.~\ref{tab:1}.
Further details on the calculations for Fig.~\ref{fig:10} and Tab.~\ref{tab:2} are provided in App.~\ref{app:E}.

\begin{table}[h!]
\small
\renewcommand{\arraystretch}{1.2}
\caption{\label{tab:2}
Comparison of error scalings $\epsilon_{\text{spec}} = c(n)t^{m}$ for circuits approximating the unitary $\exp(-\text{i} t H)$ constructed using Trotter product formulas~\cite{Suzuki1990,Suzuki1991} and via classical optimization.
The prefactors $c(n)$ (lower table) and exponents $m$ (upper table) are found by fitting to the spectral-norm distance data (see App.~\ref{app:E} for details).
Note that the $L=5$ Trotter-Suzuki product formula is constructed using the recursion relation in equation (3.14) of~\cite{Suzuki1991}.
}
\begin{ruledtabular}
\begin{tabular}{@{}lrrrrrr@{}}
\multicolumn{1}{@{}}{} &
\multicolumn{3}{c}{classically optimized} &
\multicolumn{3}{c}{Trotter}\\
\cmidrule(r{0.1em}){2-4}
\cmidrule(l{0.1em}){5-7}
L&
\multicolumn{1}{c}{n=5}&
\multicolumn{1}{c}{n=8}&
\multicolumn{1}{c}{n=10}&
\multicolumn{1}{c}{n=5}&
\multicolumn{1}{c}{n=8}&
\multicolumn{1}{c}{n=10}\\
\cmidrule(){2-2}
\cmidrule(){3-3}
\cmidrule(r{0.1em}){4-4}
\cmidrule(l{0.1em}){5-5}
\cmidrule(){6-6}
\cmidrule(){7-7}
1 & $2.9(1)$ & $2.9(1)$ & $2.9(2)$ & $1.98(4)$ & $1.97(5)$ & $1.97(5)$ \\
1 & -        & -        & -        & $2.98(4)$ & $2.98(4)$ & $2.98(4)$ \\
2 & $4.4(1)$ & $4.5(3)$ & $4.4(3)$ & -         & -         & -         \\ 
3 & $6.1(2)$ & $6.2(2)$ & $6.2(1)$ & $4.94(9)$ & $4.94(9)$ & $4.94(9)$ \\
4 & $7.0(1)$ & $6.8(2)$ & $6.6(1)$ & -        & -         & -         \\
5 & $7.9(3)$ & $8.0(5)$ & $7.9(2)$ & $4.98(2)$ & $4.98(2)$ & $4.98(2)$ \\
6 & $9(2)$   & $9.1(3)$ & $9.0(2)$ & -         & -         & -         \\
\cmidrule(l{0.1em}r{0.1em}){2-7}
1 & ${\scriptstyle8.7(7)\cdot10^{-1}}$ & ${\scriptstyle1.7(2)}$             & ${\scriptstyle2.3(3)}$             & ${\scriptstyle3.2(2)}$  & ${\scriptstyle5.33(4)}$ & ${\scriptstyle6.7(6)}$  \\
1 & -                                  & -                                  & -                                  & ${\scriptstyle2.4(1)}$  & ${\scriptstyle4.2(3)}$  & ${\scriptstyle5.4(4)}$  \\
2 & ${\scriptstyle3.7(2)\cdot10^{-2}}$ & ${\scriptstyle9.6(9)\cdot10^{-2}}$ & ${\scriptstyle9.0(1)\cdot10^{-1}}$ & -                       & -                       & -                       \\
3 & ${\scriptstyle1.8(1)\cdot10^{-3}}$ & ${\scriptstyle3.2(2)\cdot10^{-3}}$ & ${\scriptstyle4.2(1)\cdot10^{-3}}$ & ${\scriptstyle5.4(8)}$  & ${\scriptstyle10(2)}$   & ${\scriptstyle13(2)}$   \\
4 & ${\scriptstyle7.7(9)\cdot10^{-5}}$ & ${\scriptstyle1.8(2)\cdot10^{-4}}$ & ${\scriptstyle2.8(2)\cdot10^{-4}}$ & -                       & -                       & -                       \\
5 & ${\scriptstyle5(2)\cdot10^{-6}}$   & ${\scriptstyle9(5)\cdot10^{-6}}$   & ${\scriptstyle1.4(3)\cdot10^{-5}}$ & ${\scriptstyle0.36(1)}$ & ${\scriptstyle0.65(2)}$ & ${\scriptstyle0.85(3)}$ \\
6 & ${\scriptstyle3(3)\cdot10^{-7}}$   & ${\scriptstyle4(1)\cdot10^{-7}}$   & ${\scriptstyle6(2)\cdot10^{-7}}$   & -                       & -                       & -                       \\
\end{tabular}
\end{ruledtabular}
\end{table}

\section{Gate optimization methods}
\label{app:C}

In this appendix we assess the performance of the coordinatewise approach and the Newton method by considering the approximation error achieved after successive iterations of the algorithms.
For the coordinatewise method, one step of the algorithm corresponds to the update of a single variational parameter, whereas a single step of a global version of the Newton method updates all variational parameters at once.
Between these two extremes we also perform the Newton method on a gate-by-gate and layer-by-layer basis, whereby only those variational parameters within a single gate or within a single layer, respectively, are updated simultaneously while all others remain fixed.
Additionally we compare the coordinatewise and Newton method with the Adam optimizer~\cite{KiBa14}.

\begin{figure}
\centering
\includegraphics[width=0.95\columnwidth]{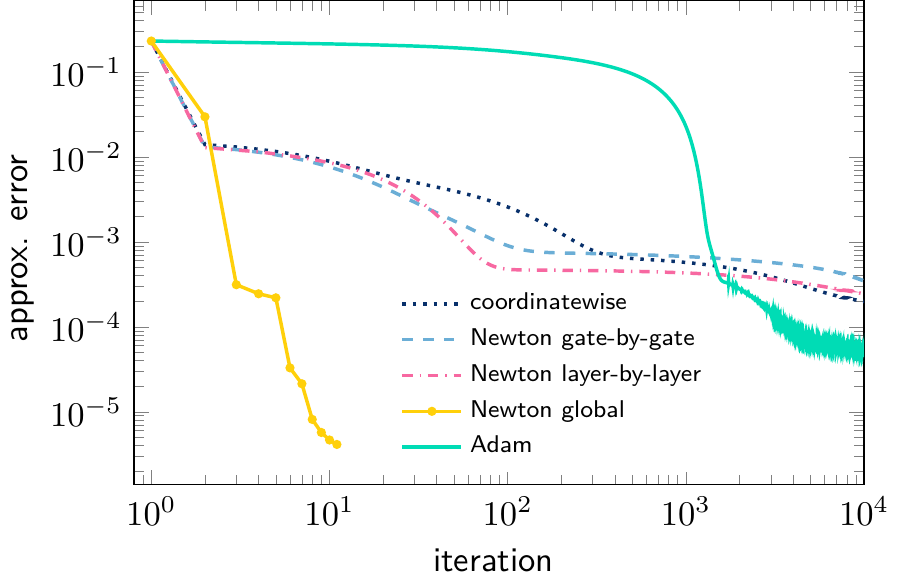}
\caption{\label{fig:11}
Approximation error of the classically optimized circuit $U(\boldsymbol{\theta})$ with respect to the exact time evolution operator $\exp(-\text{i} t H)$ achieved using various optimization algorithms.
The variational parameters $\boldsymbol{\theta}$ are initialized to zero and a single iteration corresponds to the update of all variational parameters in the circuit once.
We consider the Hamiltonian of Eq.~\eqref{eq:H} with parameters $(J,g,h) = (2.0,1.0,1.0)$, time $t = 0.05$, $n = 8$ qubits and a brickwall circuit of two layers.
}
\end{figure}

Figure~\ref{fig:11} shows the performance of each algorithm.
We observe that the global Newton method converges to the lowest approximation error in the fewest number of just ten steps.
In contrast, the Newton method on a gate-by-gate basis converges more slowly and produces a circuit with a larger approximation error, whereas the Newton method on a layer-by-layer basis produces results which lie somewhere in between the global and local update methods.
Remarkably, after ten thousand iterations, the very efficient coordinatewise update method compares well to both batched Newton methods.
Finally, after many iterations spent near local minima, the Adam optimizer with a step size $\alpha = 10^{-4}$ and exponential decay rates $\beta_{1} = 0.9$ and $\beta_{2} = 0.99$, converges to an approximation error less than $10^{-4}$.

While the global Newton method is the most accurate method, it is also the most computationally expensive since it requires explicit calculation of the Hessian matrix which, for a circuit with many parameters, becomes cumbersome to evaluate.
We have found that quasi-Newton methods such as BFGS\cite{Nocedal2009} and L-BFGS\cite{Liu1989} also give satisfactory results without having to calculate the Hessian explicitly.

In the following paragraphs we investigate under which circumstances gradients of the cost function tend to zero such that the optimization landscape of the cost function is said to be in a barren plateau~\cite{McClean2018, Cerezo2021} and the circuits therefore cannot be optimized.
In particular we investigate the magnitudes of the gradients of the normalized cost function
\begin{equation}\label{eq:SNormalizedFidelity}
 F(\boldsymbol{\theta}) = \Re \{ \tr[ U^{\dag}(\boldsymbol{\theta}) \exp(-\text{i} \tau H) U_{s-1}] \} / 2^{n}
\end{equation}
as a function of the number of qubits $n$ and the time step $\tau$.

\begin{figure}
\centering
\includegraphics[width=0.95\columnwidth]{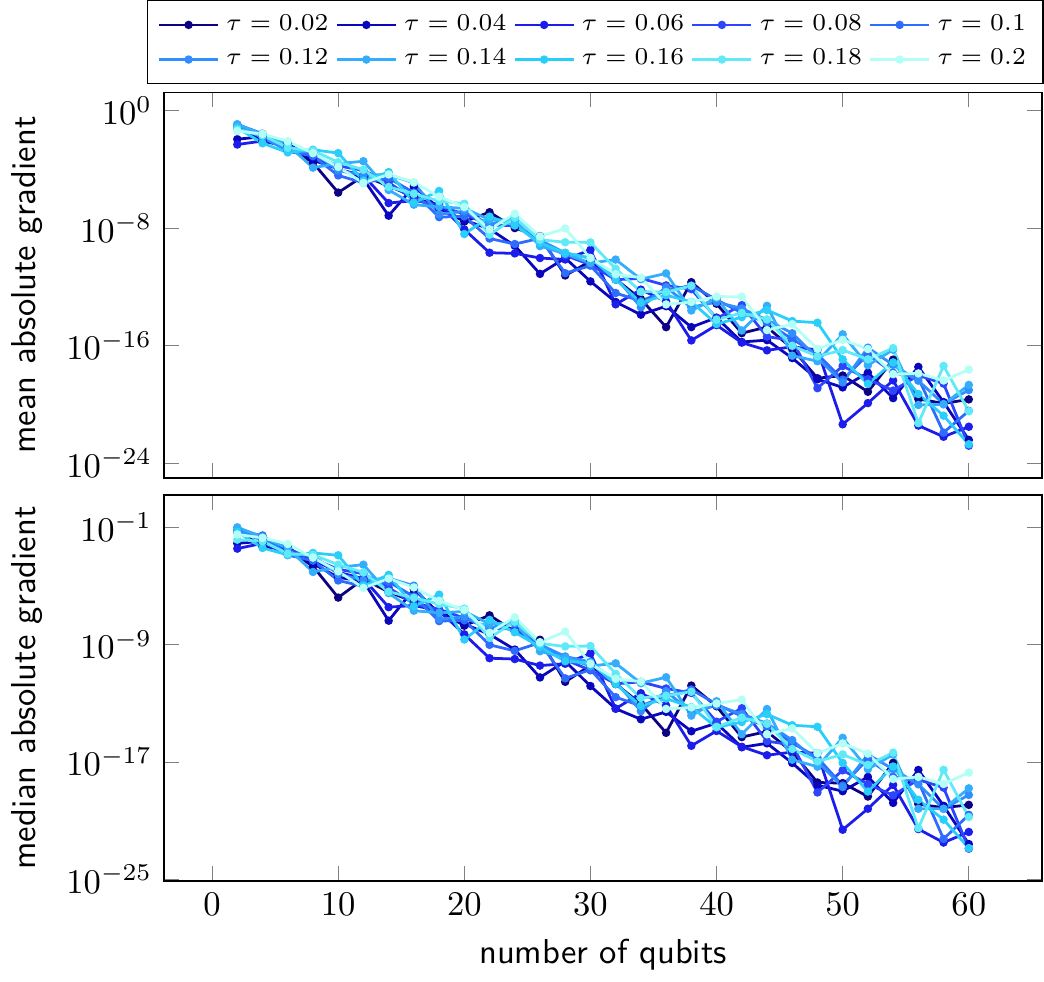}
\caption{\label{fig:12}
Mean (top) and median (bottom) absolute gradients of the first step $(s=1)$ of the sequential optimization procedure using random initialization.
Data are for system sizes of $n$ qubits and a range of Taylor time steps $\tau$.
}
\end{figure}

\begin{figure}
\centering
\includegraphics[width=0.95\columnwidth]{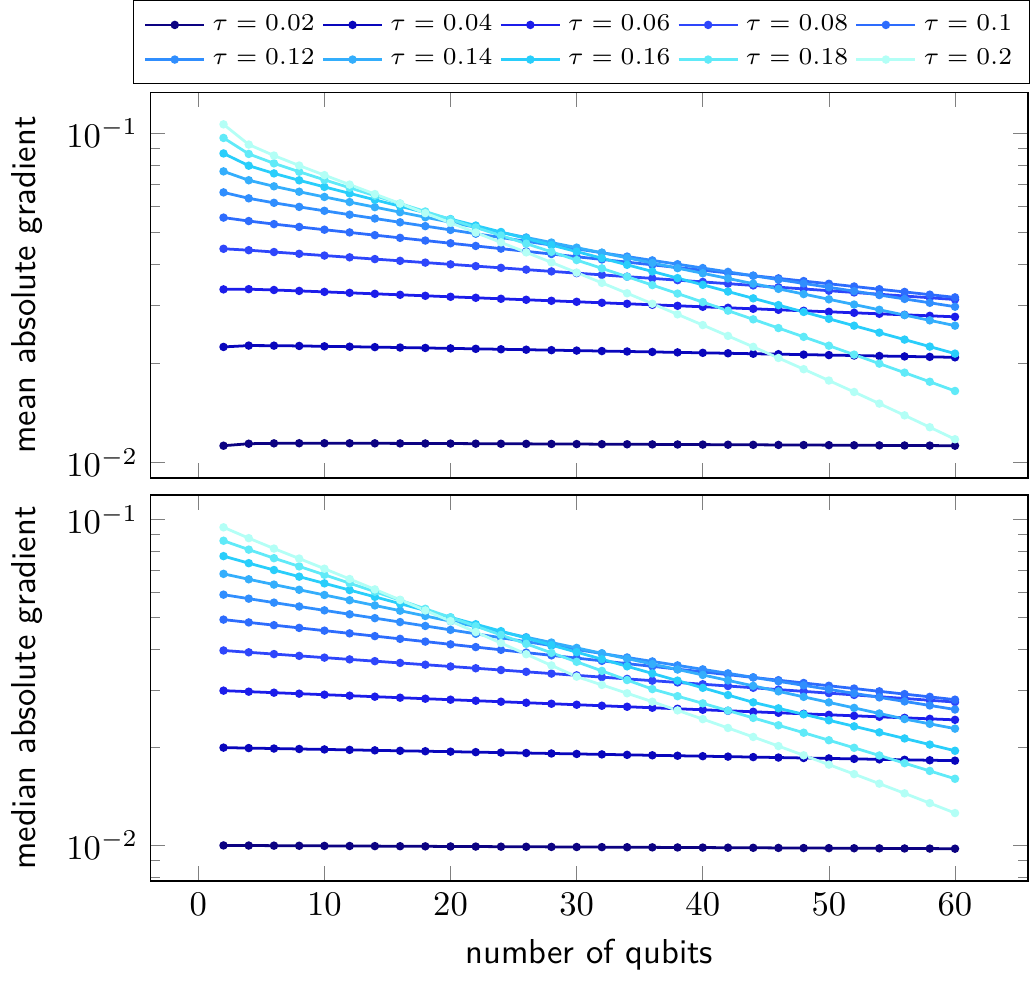}
\caption{\label{fig:13}
Mean (top) and median (bottom) absolute gradients of the first step $(s=1)$ of the sequential optimization procedure using identity initialization.
Data are for system sizes of $n$ qubits and a range of Taylor time steps $\tau$.
}
\end{figure}

To investigate the effect of different circuit initializations, we concentrate on the first step ($s=1$) of the Taylor based sequential optimization method where $U_{0}=\mathds{1}$ in the cost function Eq.~\ref{eq:SNormalizedFidelity}.
For a small time step $\tau$, the unitary evolution operator $\exp(-\text{i} \tau H)$ is approximated by a Taylor approximation and the absolute values of the gradients of the cost function $\lvert \grad_{\boldsymbol{\theta}} F(\boldsymbol{\theta}) \rvert$ are evaluated for a range of $n$ and $\tau$.
Importantly, the values of $\boldsymbol{\theta}$ are either chosen randomly, which we call random initialization (Fig.~\ref{fig:12}), or they are all set to zero such that, initially, $U(\boldsymbol{\theta})=\mathds{1}$, which we call identity initialization (Fig.~\ref{fig:13}).
We study the Ising Hamiltonian Eq.~\eqref{eq:H} with $(J,g,h)=(2.0,1.0,1.0)$.

For the random initialization in Fig.~\ref{fig:12} we see a clear exponential decay of the mean and median of the absolute values of the cost function gradients as the number of qubits increases.
This exponential decay is present for all values of the time step $\tau$ and therefore random initialization leads to barren plateaus.
For identity initialization in Fig.~\ref{fig:13} we find that both the mean and median of the absolute value of the gradients decay much more slowly as a function of $n$ and furthermore, the rate at which they decay can be reduced by decreasing the time step $\tau$.

\begin{figure}
\centering
\includegraphics[width=0.95\columnwidth]{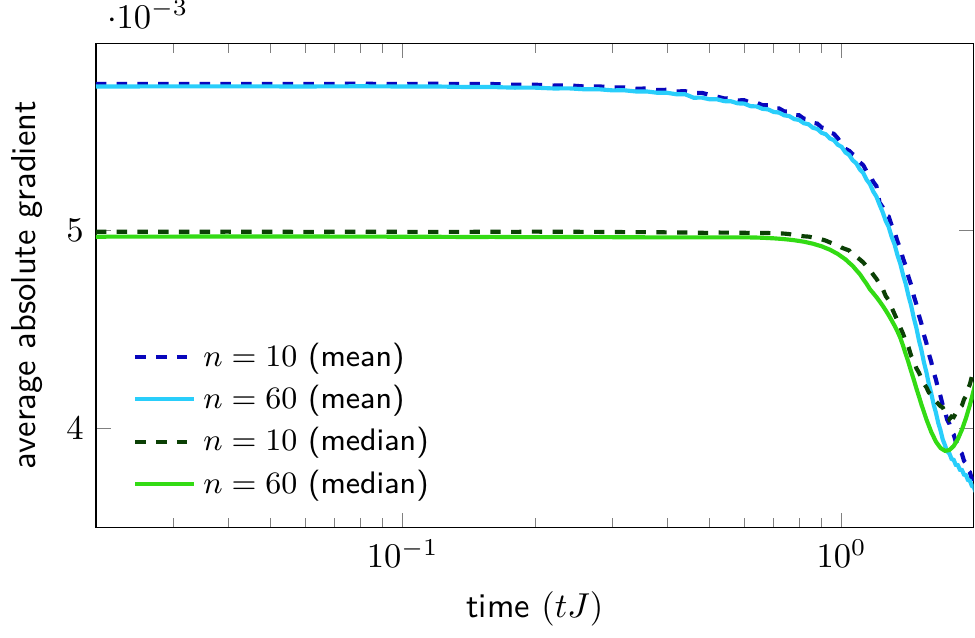}
\caption{\label{fig:14}
Average absolute gradients (mean and median) as a function of time for systems of $n=10$ and $n=60$ qubits. 
We compute the gradient of the cost function before the first classical optimization step (e.g.\ L-BFGS) at each slice $s$.
Data are presented as a function of time where $t=\tau s$.
}
\end{figure}

Next we examine whether barren plateaus emerge in subsequent steps of the sequential optimization scheme in the case where identity initialization is used in the first step.
To this end, we again examine the mean and median of the absolute values of the gradients for systems of $n=10$ and $n=60$ qubits and brickwall circuits of two layers.
More precisely, we examine $\lvert \grad_{\boldsymbol{\theta}} F(\boldsymbol{\theta}) \rvert$ at the beginning of each slice $s$ of the sequential optimization, i.e.\ before the first iteration of the classical optimizer.
We set the total number of slices in the sequential optimization to $S=200$ and the total time $tJ=4.0$ such that $\tau=t/S=0.01$.
The results shown in Fig.~\ref{fig:14} demonstrate that we do not encounter barren plateaus.
This can be understood as follows.
The identity initialization provides a good initialization for the first slice of the sequential optimization.
After optimizing the first slice, we are left with a set of optimized parameters $\boldsymbol{\theta}_{1}$ which provide a good initialization for the second slice of the sequential optimization, after which we are left with $\boldsymbol{\theta}_{2}$ and so on.

\section{5-qubit results}
\label{app:D}

\begin{figure}
\centering
\includegraphics[width=0.95\columnwidth]{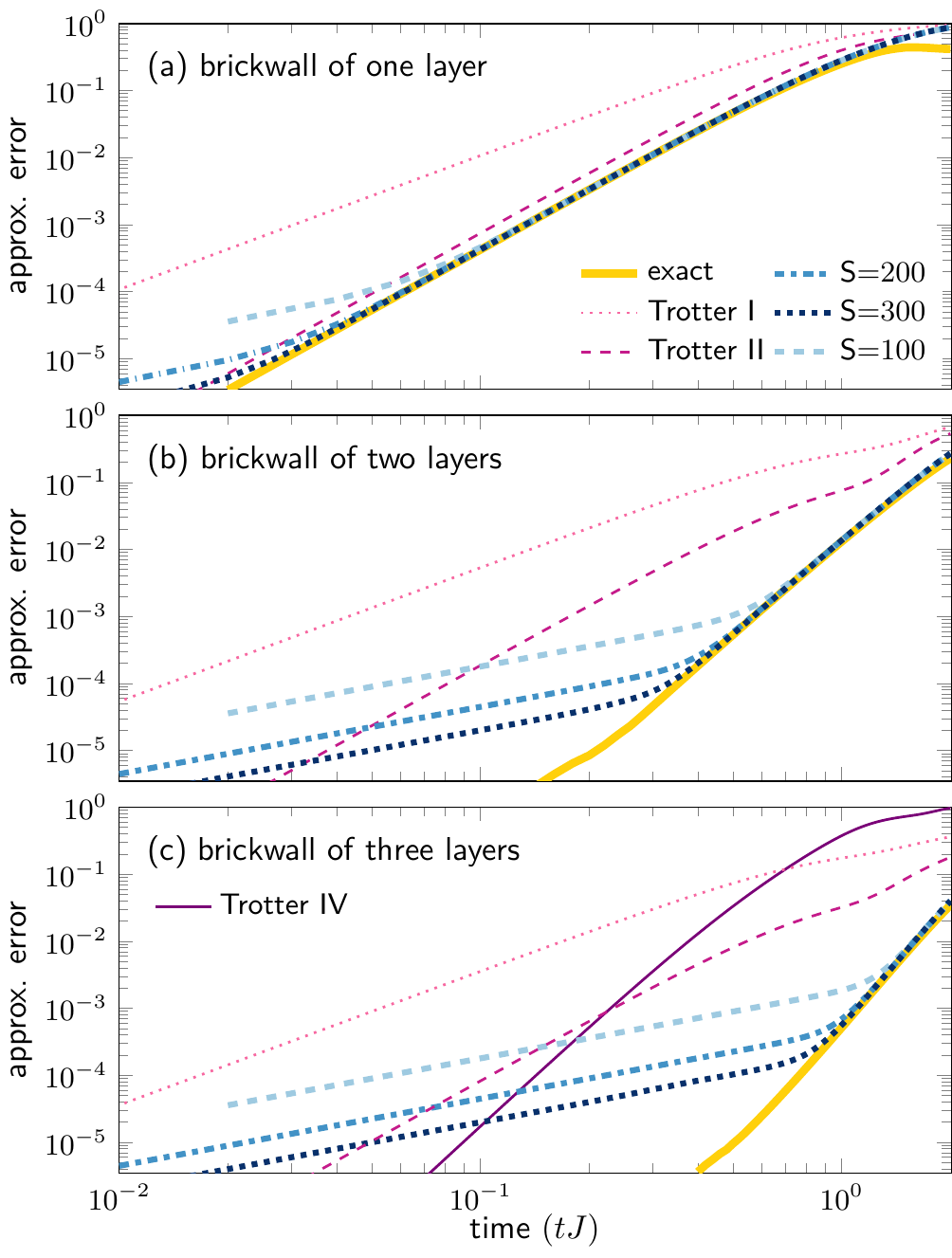}
\caption{\label{fig:15}
Classically optimized Hamiltonian simulation (thick lines) vs.\ Trotter product formulas (thin lines) for a brickwall circuit of depth (a) one, (b) two, and (c) three.
The approximation error of the approximate evolution operator $U_{\text{approx}}$ with respect to the exact one $\exp(-\text{i} t H)$ is shown as a function of time $t$.
We consider $H = 2 \sum_{k = 1}^{4} Z_{k} Z_{k+1} + \sum_{k = 1}^{5} X_{k} + \sum_{k = 1}^{5} Z_{k}$.
For the Trotter results $U_{\text{approx}}$ is a Trotter product formula of $1$st (dotted), $2$nd (dashed) and $4$th (solid) order.
For the classically optimized results $U_{\text{approx}} = U(\boldsymbol{\theta})$ is the circuit with parameters $\boldsymbol{\theta}$ after either the \textsf{exact} (thick solid) or Taylor based (thick dashed) optimization procedures as outlined in Sec.~\ref{sec:Results}.
}
\end{figure}

In this appendix we supplement the data shown in Fig.~\ref{fig:1}, which is calculated for a system of $n = 8$ qubits, with the equivalent data calculated for $n = 5$ qubits.
Figure~\ref{fig:15} shows the approximation error as a function of time for brickwall circuits of one (a), two (b) and three (c) layers. 

\section{Error scaling}
\label{app:E}

In this appendix we provide further details on the calculation of the data in Figs.~\ref{fig:3} and~\ref{fig:10} and on the fitting procedure used to find the prefactors and exponents given in Tabs.~\ref{tab:1} and~\ref{tab:2}.

The results in Figs.~\ref{fig:3} and~\ref{fig:10} are calculated using the \textsf{exact} method and a quasi-Newton optimization scheme.
For each value $(n,L)$ we optimize $S = 40$ slices in sequence and for each slice we demand that the two-norm of the vector of gradients falls below $10^{-6}$ to achieve convergence.
The PastaQ.jl~\cite{pastaq} packages were used to calculate the cost function and its gradients.

The choice of scaling function $\epsilon = c(n)t^m$ is motivated by the empirical observation that the data lie on a straight line in the log-log scale.
This choice is further supported by the known error scaling for Trotter product formulas.
To estimate the exponent $m$ from the data we perform a least-squares fit of $y(x) = \log(c) + m x$ where $x = \log(t)$ and $y = \log(\epsilon(t))$, time is rescaled by $t \to tJ$ and $\epsilon(t)$ is either $\epsilon_{\text{approx}}$ of Eq.~\eqref{eq:ApproxError} or $\epsilon_{\text{spec}}$ of Eq.~\eqref{eq:SSpecDist}.
The fit is performed using a subset of $\kappa$ data points between an initial time $t_{\text{i}}$ and a final time $t_{\text{f}}$ chosen such that those data are approximately linear on a log-log scale; below $t_{\text{i}}$ the data suffer from the finite precision of the optimization while data above $t_{\text{f}}$ have large errors causing a deviation from the power-law scaling.
Multiple fits are performed using the $\kappa$ data points.
First we take each pair of $l = 2$ consecutive data points and fit the power law using a least-squares routine.
We then repeat this for $l \in [3, 4, \ldots \kappa]$ such that we are left with a set of prefactors $\{c_j\}$ and exponents $\{m_j\}$.
The mean values of these sets are calculated to give the $c$ and $m$ quoted in Tabs.~\ref{tab:1} and~\ref{tab:2}.
The uncertainties of the prefactors and exponents are estimated by the maximum deviation from the mean value.

\section{Experimental noise}
\label{app:F}

To assess how experimental noise affects our results, in this appendix we consider the effect of gate errors as they are present in current quantum devices.

\begin{figure}
\centering
\includegraphics[width=0.95\columnwidth]{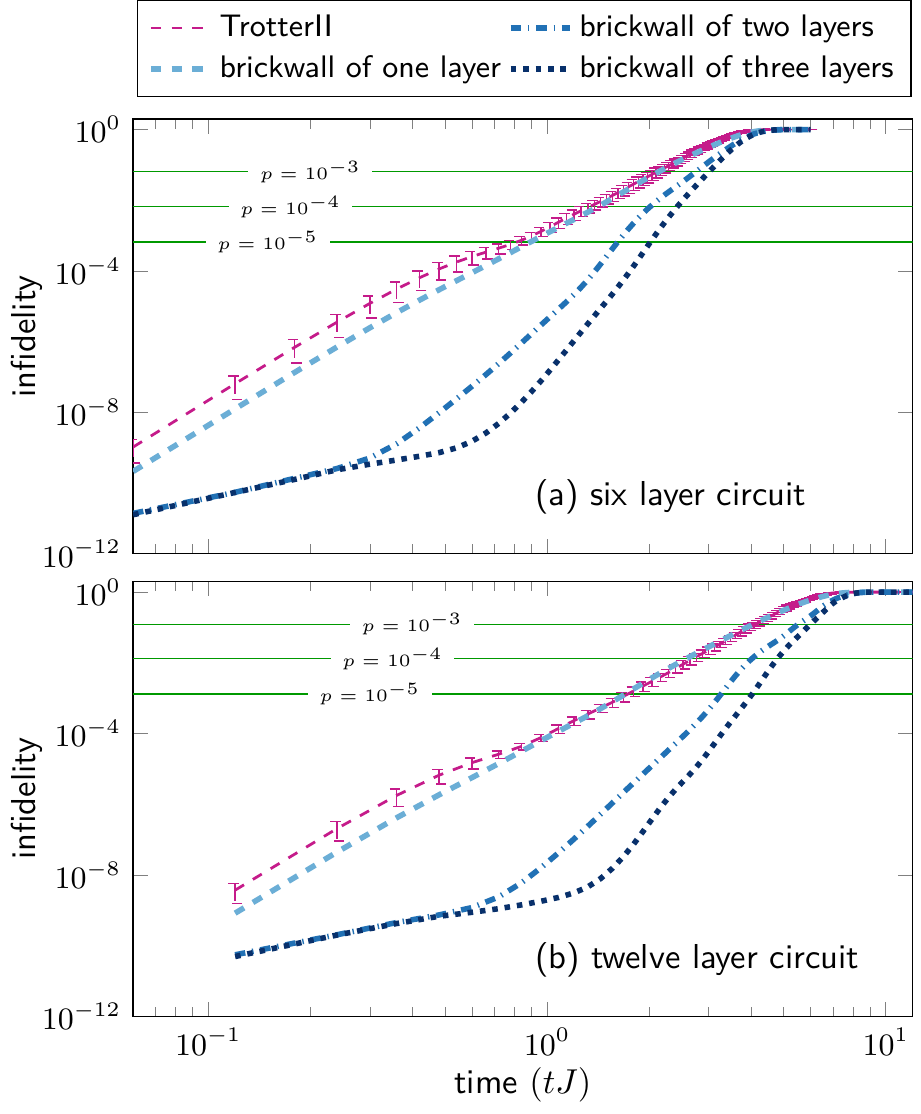}
\caption{\label{fig:16}
Time-evolved $n = 12$ qubit state infidelity averaged over $50$ random initial bitstring product states plotted as a function of time for (a) circuits of six layers and (b) circuits of twelve layers.
We consider the Hamiltonian of Eq.~\eqref{eq:H} with parameters $(J,g,h) = (1.0,1.0,1.0)$.
The time evolution circuits approximating $\exp(-\text{i} t H)$ are constructed by appending multiple second-order Trotter circuits (thin dashed line) or multiple classically optimized circuits (thick lines) as explained in the text.
The solid horizontal lines approximate the total infidelity due to noisy two-qubit gates for various two-qubit gate error rates $p$.
The error bars on the data are negligibly small compared to the scale of the figure and are therefore omitted for clarity.
}
\end{figure}

In Fig.~\ref{fig:16} we show the infidelity $1-\lvert\braket{\psi_\text{exact}}{\psi_\text{approx}}\rvert^2$ as a function of time between states evolved using the exact unitary operator $\ket{\psi}_\text{exact} = \exp(-\text{i} t H)\ket{\psi}_\text{init}$ and those evolved using approximating circuits $\ket{\psi}_\text{approx} = U_{\text{approx}} \ket{\psi}_\text{init}$.
The infidelity is averaged over a set of $50$ initial states $\ket{\psi}_\text{init}$ which are drawn from the set of random bitstrings.

The approximating circuits $U_\text{approx}$ are constructed by appending multiple circuits found using a second-order Trotter product formula or via classical optimization.
More specifically, the six layer circuits in Fig.~\ref{fig:16} (a) are composed by appending either six \textsf{Trotter II} layers, six \textsf{brickwall of one layer} circuits, three \textsf{brickwall of two layers} circuits or two \textsf{brickwall of three layers} circuits such that all six layer circuits have the same number of two-qubit gates.
In a similar way we construct the twelve layer circuits in Fig.~\ref{fig:16} (b).

The infidelity as a function of time for the classically optimized circuits is similar in character to the approximation error, cf.\ Fig.~\ref{fig:1}.
At early times there is regime where the infidelity increases with a shallow slope due to the Taylor approximation while at larger times the infidelity increases with a larger slope related to the expressiveness of the ansatz.
As is also the case for the approximation error, classically optimized circuits can achieve infidelities which are orders of magnitude smaller than Trotter product formulas of the same two-qubit gate count.

We estimate the effect of noise by approximating the total error arising from two-qubit gates.
To the lowest-order approximation, the fidelity due to noisy two-qubit gates is given by $(1-p)^{K}$ where $p$ is the average error rate of two-qubit gates and $K$ is the number of those gates in the circuit.
In Fig.~\ref{fig:16} the approximate infidelity due to noisy two-qubit gates is indicated by a horizontal line for various values of $p$.
The minimum infidelity achievable is set by the noise and therefore the improvement obtained by using classically optimized circuits instead of Trotter product formulas is also limited by noise.
We see in Fig.~\ref{fig:16} that, as $p$ decreases, i.e.\ hardware gate fidelities improve, the advantage of using classically optimized Hamiltonian simulation over Trotter product formulas increases.

\bibliography{bibliography}

\end{document}